\documentclass[review]{elsarticle}

\usepackage{hyperref}
\usepackage{amsmath}
\usepackage{bm}

\journal{Computers and Geotechnics}

\begin{document}

\begin{frontmatter}

\title{A comprehensive study of coupled LBM-DEM with immersed moving boundary}


\author[mymainaddress]{G. C. Yang}

\author[mymainaddress]{L. Jing}

\author[mymainaddress]{C. Y. Kwok
\corref{mycorrespondingauthor}}
\cortext[mycorrespondingauthor]{Corresponding author}
\ead{fkwok8@hku.hk}

\author[mysecondaryaddress]{Y. D. Sobral}

\address[mymainaddress]{Department of Civil Engineering, The University of Hong Kong, Haking Wong Building, Pokfulam Road, Hong Kong}
\address[mysecondaryaddress]{Departamento de Matem\'{a}tica, Universidade de Bras\'{i}lia, Campus Universit\'{a}rio Darcy Ribeiro, 70910-900 Bras\'{i}lia, DF, Brazil}

\begin{abstract}
A systematic study is carried out on a fully resolved fluid-particle model which couples the Lattice Boltzmann Method (LBM) and the Discrete Element Method (DEM) using an immersed moving boundary technique. Similar algorithms have been reported in the past decade, however, the roles of major model parameters are yet to be fully understood. To examine various numerical errors, a series of benchmark cases with a wide range of Reynolds number are performed, starting from a single stationary particle to multiple moving particles. It is found that for flow with low and intermediate Reynolds numbers, 20 fluid cells per one particle diameter are necessary to achieve sufficient accuracy (within 5\%). For a flow with high Reynolds number, a turbulence model shall be incorporated so that the effects of unresolved small eddies can be captured in an accurate and efficient manner. Besides, the LBM-DEM results are also sensitive to the relaxation time, especially when the spatial resolution is inadequate. A large relaxation time can introduce additional diffusion of fluid momentum into the fluid-particle system, leading to weakened hydrodynamic interactions. By choosing a small relaxation time greater than the lower limit 0.5, a small fluid compressibility error and a strong coupling between fluids and particles can be achieved, at the cost of computational effort. The test cases also demonstrate the capability of LBM-DEM to describe the rheology of particle suspensions by capturing the pore-scale hydrodynamic interactions. Finally, a guideline for quickly establishing a high-quality LBM-DEM model is provided.
\end{abstract}

\begin{keyword}
Lattice Boltzmann Method \sep Discrete Element Method \sep LBM-DEM \sep immersed moving boundary \sep fluid-particle interaction \sep granular collapse
\end{keyword}

\end{frontmatter}


\section{Introduction}
\label{sec:introduction}
Particle movements in fluid flows are commonly encountered in nature and industry, characterized by the complex fluid-particle interactions. For example, the generation of excess pore fluid pressure is found to be responsible for the fast or slow dynamics of granular materials sliding down on a slope \cite{Iverson2000}. Another example in petroleum industry is the unfavorable sand production from wells in weakly-bonded rock matrix caused the induced large fluid drag force \cite{Vardoulakis1996}. The strong coupling effect between fluids and particles has an vital role underlying these micro-behaviors. Benefiting from the highly advanced computer technology, numerical simulations have become appealing tools to study the fluid-particle interaction problems, but it requires an accurate description of momentum exchange at the fluid-particle interface.

Classic laws governing the fluid-particle interactions have been developed and verified long time ago based on massive physical experiments and rigorous theoretical analysis, such as the well-known Darcy's law \cite{Whitaker1986} and the Ergun equation \cite{Ergun1952, Macdonald1979}. These semi-empirical relations usually serve as the underlying assumptions for successful multiphase numerical simulations, and for instance, the coupling between Computational Fluid Dynamics (CFD) and Discrete Element Method (DEM) \cite{Jing2016}. In this kind of CFD-DEM technique, a fluid cell always has a size larger than the particles, resulting in an averaged porosity that governs the fluid dynamics and the resultant hydrodynamics forces. The coupling method with fluid cells larger than particles is denoted as the coarse-grid method in this work. Although the coarse-grid method has been successfully applied for a variety of problems \cite{Jing2016, Zhao2013, Kloss2012}, it can only provide limited pore-scale information. Therefore, fine-grid CFD-DEM method, in which one particle covers multiple fluid cells, has also been developed to achieve a more general and accurate description of fluid-particle interaction via an Immersed Boundary Method \cite{Kim2001, Peskin2002, Hager2013}, instead of heavily relying on the semi-empirical relations. Nevertheless, the fine-grid CFD-DEM method is sometimes numerically prohibitive due to its high computational demand of mesh generation and solving the nonlinear Navier-Stokes equation, even for a small system involving several hundreds of particles \cite{Hager2013}.

Alternatively, the Lattice Boltzmann Method (LBM) can be applied, in place of the conventional CFD, together with DEM, for the simulation of fluid-particle interaction problems. LBM is an approximation of the incompressible Navier-Stokes equation at the mesoscopic scale based on kinetic theory \cite{Qian1992, Chen1998}. When LBM is coupled with DEM, it shares the advantages of fine-grid CFD-DEM method, referring to the fully resolved pore-scale fluid flows and explicit calculation of hydrodynamic forces, and it is more efficient than CFD-DEM, benefiting from the much simplified governing equations with an excellent parallel computing performance \cite{Chen1998}. By using LBM, no iteration is required to solve for the fluid velocity field and pressure field. Besides, the nature of LBM, which is a description of the whole fluid system by a collection of molecules, facilitates an intrinsic coupling between fluids and solid particles. As a result, reasonably large-scale problems can be simulated by the LBM-DEM technique accurately and efficiently.

For the coupling between LBM and DEM, there are two commonly used approaches, namely, the Momentum Exchange (ME) method \cite{Ladd1994, Aidun1998, Cheng2018} and the Immersed Moving Boundary (IMB) method \cite{Noble1998, Holdych2003}. Conceptually, the ME and IMB methods share the same basic idea by enforcing the no-slip boundary condition between fluids and solids with a hydrodynamic interaction according to the conservation law of momentum. In the ME method, the no-slip boundary condition is achieved via a bounce-back scheme at the fluid-particle interface \cite{Ladd1994}. Whereas, in the IMB method, the LBM collision operator is modified by following the non-equilibrium bounce-back principle if a fluid cell is covered by solids \cite{Noble1998}. Several studies in literature aimed at directly comparing the ME and IMB methods \cite{Han2011, Rettinger2017}. In our work, the IMB method is chosen to couple LBM with DEM, because it has a better sub-grid scale (SGS) resolution and thereby, fewer fluctuations are observed on the solid geometry and the hydrodynamic forces as particles move across fluid cells. It is also worth mentioning that we are aware of that improvements of the ME method have been developed by introducing a more complicated interpolated bounce-back scheme \cite{Rettinger2017} to improve the overall accuracy. However, the interpolation requires to access additional information at the neighboring fluid cells. As a result, the locality of the LBM calculation is lost. Besides, the interpolation may also cause unexpected numerical issues when particles are in close vicinity.

Despite the fact that successful applications of the coupled LBM-DEM model via an IMB technique have been reported across multiple disciplines \cite{Han2007, Owen2011, Lomine2013, Han2013, Leonardi2016}, there is still a lack of a systematic study about the accuracy, stability and efficiency of the LBM-IMB-DEM approach and about the role of various parameters used in these numerical simulations. In the current computational practice, a spatial resolution of 10 fluid cells per one particle diameter is commonly used \cite{Han2011}. However, more recently, Rettinger simulated a single particle settling in an ambient fluid \cite{Rettinger2017}, and it was reported that a higher resolution of 24 fluid cells per one particle diameter is required so that the particle velocity error is below 5\%. And to maintain the same accuracy, the resolution needs to be further increased as the Reynolds number of the flow increases. The determination of a sufficient spatial resolution is essential for LBM-DEM simulations since it is the key parameter affecting the simulation time. Different from the spatial resolution, we find that the role of the other important model parameter, referring to the LBM relaxation time, is less discussed. It is well known that the relaxation time can affect the accuracy and stability of LBM significantly \cite{Luo2011}, so its tremendous influence on the performance of the LBM-DEM model might also be expected.

The goal of this study is to provide a quantitative investigation on the important roles of the related parameters in the three-dimensional (3D) LBM-DEM model, including the lattice resolution ($N$), the solid ratio ($\varepsilon$), the relaxation time ($\tau$), as well as the LBM and DEM time steps ($\delta_t$ and $\Delta t$). Apart from the influence of the adopted model parameters, additional numerical errors can be caused by particle motions and by particle-fluid-particle interactions. Therefore, in order to consider various sources of errors, LBM-DEM simulation of different fluid-particle interaction problems involving a single particle or multiple particles which can be immobile and movable are carried out.

The rest of this paper is organized as following: Section \ref{sec:method} introduces the numerical methods, followed by the coupling scheme. Then, four benchmark cases are studied following the order of increasing complexity. The roles of the lattice resolution and the relaxation time are first investigated in detail via a simple problem of Poiseuille flow past a fixed sphere in Section~\ref{subsec:poiseFlow}. Section~\ref{subsec:particleSettle} presents the simulation of a heavy particle settling in an ambient fluid to test the sub-cycling scheme and to examine the numerical error caused by particle moving across multiple fluid cells. And then, the LBM-DEM model for densely packed particle systems is well validated by simulating a one-dimensional flow through a porous medium driven by various pressure differences in Section~\ref{subsec:flowPack}. The necessity of incorporating a turbulence model at high Reynolds number is demonstrated. After that, the capability of capturing the complex particle-fluid-particle interactions is highlighted in Section~\ref{subsec:couette}, in which a concentrated suspension in a planar Couette flow is simulated. Finally, a guideline based on our findings to efficiently establish a successful LBM-DEM model is presented in Section~\ref{sec:conclusions}.

\section{LBM-DEM formulation}
\label{sec:method}
\begin{figure}[bt]
\centering
\includegraphics{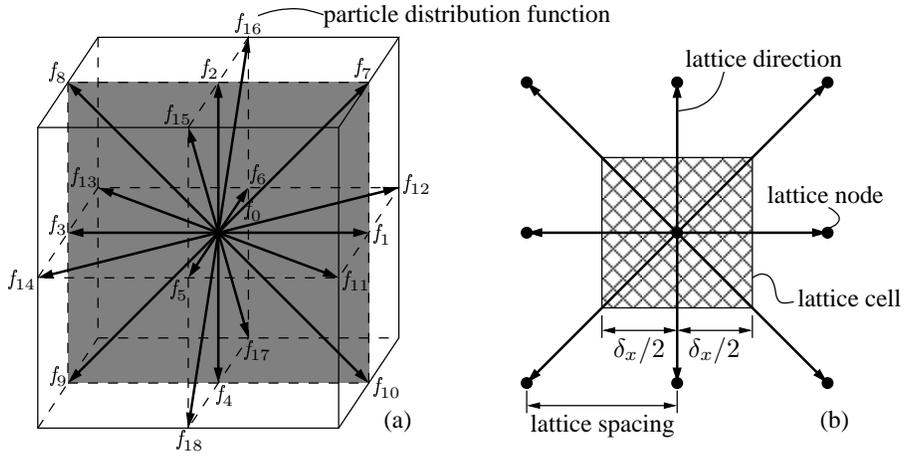}
\caption{(a) A D3Q19 lattice structure for 3D LBM simulations with 19 lattice velocities. Each lattice velocity is associated with a PDF from $f_0$ to $f_{18}$; the shaded plane is redrawn in (b) showing the lattice direction, lattice node, lattice spacing and a two-dimensional projection of a cubic lattice cell.}
\label{fig:lattice}
\end{figure}

LBM solves the hydrodynamics based on the kinetic theory in a mesoscopic scale \cite{Chen1998}. The whole fluid system is described by a collection of molecules residing on a regular Cartesian mesh (lattice) with cubic cells. The number of fluid molecules at each lattice node is quantified by a set of particle distribution functions (PDFs) with pre-defined discrete directions pointing to the neighboring lattice nodes. The PDF at time $t$ positioned at $\textbf{x}$ pointing to the $i$th direction is denoted as $f_i(\textbf{x}, t)$. In this study, a D3Q19 lattice structure \cite{Qian1992}, as shown in Fig.~\ref{fig:lattice}(a), is used for 3D LBM simulations. 19 discrete velocities are used, instead of 15 or 27, to achieve a good balance between accuracy and efficiency. The definitions of lattice direction, lattice node, lattice cell and lattice spacing ($\delta_x$) are illustrated in Fig.~\ref{fig:lattice}(b).

Different from the conventional CFD that solves the nonlinear partial differential equations in terms of macroscopic variables, such as the fluid density $\rho_f$ and the fluid velocity $\textbf{u}_f$, the governing equation in LBM describes the evolution of PDFs. With a BGK approximation (named after Bhatnagar, Gross, and Krook \cite{Bhatnagar1954}), the governing equation is written as:

\begin{equation}
f_i \left( \mathbf{x}+\mathbf{c}_i\delta_t, t+\delta_t \right) - f_i \left( \mathbf{x}, t \right) = -\frac{1}{\tau} \left[ f_i \left( \mathbf{x}, t \right) - f_i^{eq} \left( \mathbf{x}, t \right) \right].
\label{eq:lbe}
\end{equation}

The left-hand side (LHS) of Eq.~\eqref{eq:lbe} is the \textit{streaming} process, during which the PDFs are passed to the neighboring lattice nodes (from $\mathbf{x}$ to $\mathbf{x}+\mathbf{c}_i\delta_t$) with a lattice velocity $\mathbf{c}_i$ along the $i$-th direction over a LBM time step $\delta_t$. The right-hand side (RHS) of Eq.~\eqref{eq:lbe} is the \textit{collision} process, during which the PDFs are linearly relaxed towards the equilibrium distribution functions (EDFs), $f_i^{eq}(\mathbf{x}, t$), with a single relaxation time $\tau$. The EDF adopted here is the Maxwellian one, which can be expanded into a Taylor series with respect to the macroscopic fluid velocity $\mathbf{u}_f$, as \cite{Aidun2010}:

\begin{equation}
f_i^{eq} = w_i \rho_f \left[ 1+\frac{\mathbf{c}_i \cdot \mathbf{u}_f}{c_s^2}+\frac{\left(\mathbf{c}_i \cdot \mathbf{u}_f\right)^2}{2c_s^4}-\frac{u_f^2}{2c_s^2} \right],
\label{eq:feq}
\end{equation}
where $w_i$ is the weight associated with the lattice velocity $\mathbf{c}_i$, whose values are summarized in Table~\ref{tab:weight}. The speed of sound $c_s$ for D3Q19 is $1/\sqrt{3}$ in lattice units \cite{Aidun2010}. The ratio between the magnitude of fluid velocity and the speed of sound is defined as the Mach number, that is $M = u_f/c_s$.

\begin{table}[bt]
\caption{Summary of the weight $w_i$ for PDF $f_i$ with lattice velocity $\mathbf{c}_i$. Note that the summation of all the weights shall be equal to the unity.}
\begin{tabular}{l l l}
\hline
PDFs, $f_i$ & Lattice velocity, $\mathbf{c}_i$ & Weight, $w_i$\\
\hline
$f_0$            & (0, 0, 0)                                                     & 1/3\\
$f_1$ - $f_6$    & ($\pm$1, 0, 0), (0, $\pm$1, 0), (0, 0, $\pm$1)                & 1/18\\
$f_7$ - $f_{18}$ & ($\pm$1, $\pm$1, 0), ($\pm$1, 0, $\pm$1), (0, $\pm$1, $\pm$1) & 1/36\\
\hline
\end{tabular}
\label{tab:weight}
\end{table}

Based on the fundamental laws of mass and momentum conservations, the macroscopic fluid density $\rho_f$ and velocity $\mathbf{u}_f$ can be reconstructed from the zeroth-order and first-order velocity moments of the PDFs, as:

\begin{equation}
\rho_f = \sum \limits_{i=0}^{18} f_i,
\label{eq:density}
\end{equation}

\begin{equation}
\rho_f \textbf{u} = \sum \limits_{i=0}^{18} \textbf{u}_i f_i.
\label{eq:velocity}
\end{equation}

The Navier-Stokes equation can be recovered from Eq.~\eqref{eq:lbe} via a multi-scale (Chapman-Enskog) expansion \cite{He1997}, and a relationship between the relaxation time $\tau$, the LBM time step $\delta_t$, the lattice spacing $\delta_x$ and the kinematic fluid viscosity $\nu_f$ is obtained as:

\begin{equation}
\nu_f = c_s^2 \left(\tau-\frac{1}{2}\right) \frac{\delta_x^2}{\delta_t}.
\label{eq:viscosity}
\end{equation}

In Eq.~\eqref{eq:viscosity}, $c_s$ and $\tau$ are model constants and $\nu_f$ is the material property. The LBM time step $\delta_t$ is dependent on the used discretization of the lattice grid with spacing $\delta_x$. The other macroscopic variable, pressure $p$, can be calculated from the fluid density by the equation of state \cite{He1997}:

\begin{equation}
p = c_s^2 \rho_f.
\label{eq:pressure}
\end{equation}

The major source of compressibility error in LBM is the truncated Taylor expansion of the EDFs when the higher order terms of the Mach number are dropped off. To approximate an incompressible flow, it must fulfill $M \ll 1$. The incompressible requirement in LBM simulations poses a constraint on the LBM time step and thereby affects the time step for the particle simulations. The synchronization issue between LBM and DEM will be addressed in Section~\ref{subsec:coupling}.

\subsection{Discrete Element Method}
\label{subsec:dem}
In many coupled fluid-particle simulations, such as fluidized bed, the inter-particle interaction is often approximately treated as an averaged lubrication force, and the motion of the particles are also treated regarding averages \cite{Davis1986}. However, for the cases where the particles are densely packed and subjected to large displacements, such as the immersed granular column collapse case \cite{Topin2012}, the interaction between contacting particles has to be accurately calculated. A thoughtful choice is to adopt DEM \cite{Cundall1979} in order to better resolve the inter-particle interactions.

\begin{figure}[bt]
\centering
\includegraphics{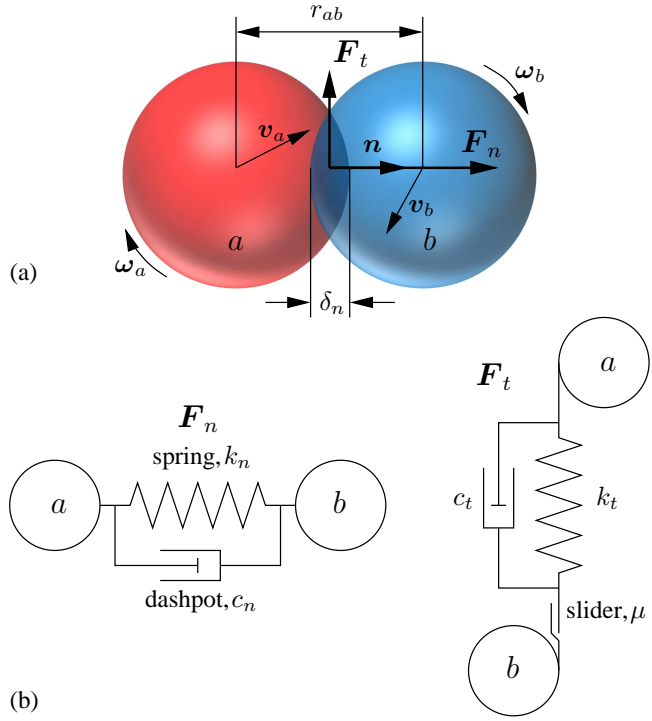}
\caption{(a) Sketch of two particles in contact: particle $a$ in red and particle $b$ in blue; (b) Schematic sketches of the spring-dashpot model for the calculation of normal force $\mathbf{F}_n$ and tangential force $\mathbf{F}_t$.}
\label{fig:dem}
\end{figure}

For the classic formulation of DEM, individual particles are taken as "rigid" bodies with "soft" contacts, allowing small overlaps between contacting objects. Fig.~\ref{fig:dem}(a) shows a contact pair between particle $a$ (in red) and particle $b$ (in blue) with a overlap equal to $\bm{\delta}_n$, which can be calculated by:

\begin{equation}
\bm{\delta}_n = (r_a+r_b-r_{ab}) \cdot \mathbf{n},
\label{eq:overlap}
\end{equation}
where $r_a$ and $r_a$ are the radii of particle $a$ and $b$, respectively. The spacing between the particle centers is denoted as $r_{ab}$. $\mathbf{n}$ is the unit normal pointing to the particle center. For simplicity, all the particles in this study are spherical.

The contact forces can be calculated based on a simple spring-dashpot model \cite{Cundall1979}, as shown in Fig.~\ref{fig:dem}(b). The normal contact force $\mathbf{F}_n$ is given by \cite{Cundall1979}:

\begin{equation}
\mathbf{F}_n = k_n \bm{\delta}_n+c_n \Delta \mathbf{u}_n,
\label{eq:normalForce}
\end{equation}
where $k_n$ and $c_n$ are the stiffness and damping coefficient in the normal direction. The relative normal velocity is denoted as $\Delta \mathbf{u}_n$. The tangential contact force $\mathbf{F}_t$ is written as \cite{Cundall1979}:

\begin{equation}
\mathbf{F}_t = k_t \int \limits_{t_{c,0}}^{t_c} \Delta \mathbf{u}_t \mathrm{d}t +c_t\Delta \mathbf{u}_t,
\label{eq:tangentForce}
\end{equation}
where $k_t$ and $c_t$ are the stiffness and damping coefficient in the tangential direction, the relative tangential velocity is denoted as $\Delta \textbf{u}_t$. The integral corresponds to an incremental spring that stores energy from the relative tangential motion, representing the elastic deformation of the particle surface since contact from time $t_{c,0}$ to $t_c$. The tangential force points to a direction opposite to the tangential displacement. Besides, the magnitude of the tangential force is limited by the Coulomb friction $\mu F_n$, at which the two contacting particles start to slide against each other. $\mu$ is the smaller of the friction coefficients of the two particles in contact.

By changing $k_n$, $k_t$, $c_n$ and $c_t$ as a function of overlap and relative velocities, different contact models (or force-displacement laws) can be proposed for the calculation of contact force ($\mathbf{F}_c$ = $\mathbf{F}_n$+$\mathbf{F}_t$). In this study, the simplified and well verified Hertz-Mindlin contact model is adopted \cite{Renzo2004}. Considering the forces (contact $\mathbf{F}_c$, gravity $\mathbf{G}$, fluid drag $\mathbf{F}_f$) and torques (contact $\mathbf{T}_c$, fluid drag $\mathbf{T}_f$) acting on a particle, its linear and angular velocities can be updated according to the Newton's second law of motion:

\begin{equation}
m\mathbf{a} = \mathbf{F}_c+\mathbf{G}+\mathbf{F}_f,
\label{eq:linearVelocity}
\end{equation}

\begin{equation}
I\dot{\bm{\omega}} = \mathbf{T}_c+\mathbf{T}_f,
\label{eq:angularVelocity}
\end{equation}
where $m$ and $I$ are the mass and moment of inertia of the particle. The translational acceleration and angular velocity are denoted as $\mathbf{a}$ and $\bm{\omega}$, respectively. The updated particle position and orientation can be calculated by taking the time integral of Eq.~\eqref{eq:linearVelocity} and Eq.~\eqref{eq:angularVelocity} via the Verlet method \cite{Verlet1967}.

\subsection{Immersed moving boundary method}
\label{subsec:imb}
\begin{figure}[bt]
\centering
\includegraphics{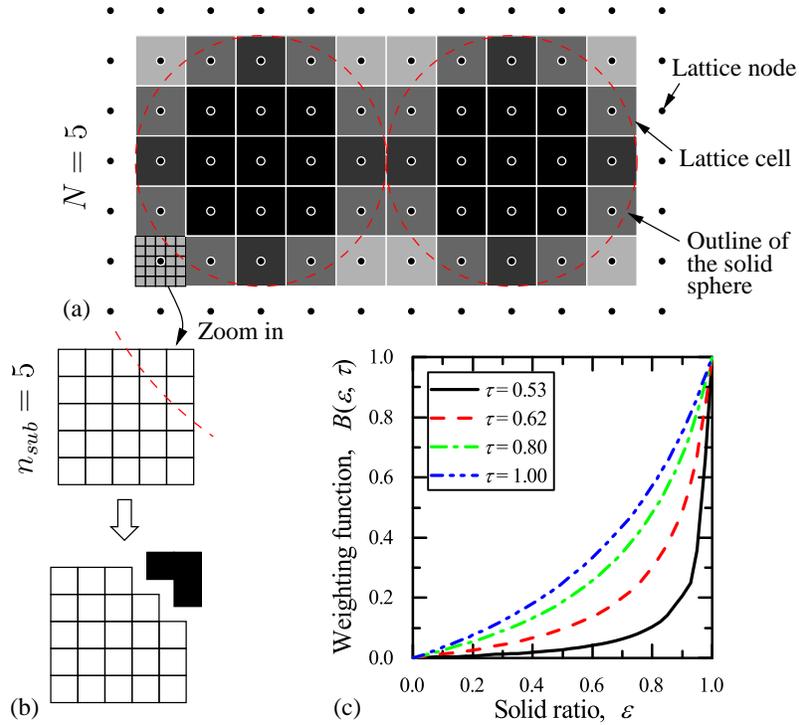}
\caption{(a) Two-dimensional sketch of two DEM spheres mapping on the LBM lattice grid with the lattice resolution $N =$ 5. The darkness of a lattice cell corresponds to its solid ratio $\varepsilon$: white ($\varepsilon$ = 0), grey (0 $<$ $\varepsilon$ $<$ 1) and black ($\varepsilon$ = 1) refer to fluid, partially saturated and solid cells, respectively; (b) One of the partially saturated cells is zoomed in and its solid ratio $\varepsilon$ is calculated via a cell decomposition method with 5 sub-slices; (c) Plot of the weighting function $B$ against the solid ratio $\varepsilon$ at various relaxation times $\tau$.}
\label{fig:SR}
\end{figure}

In this study, the adoption of LBM and DEM to simulate the fluid and particle phases, separately, necessitates an efficient and accurate coupling framework. Fig.~\ref{fig:SR}(a) shows a two-dimensional sketch of two DEM spheres mapping on the LBM lattice grid. In this sketch, the lattice resolution, $N$, defined as the number of lattice cells per one particle diameter, is equal to 5. The darkness of the lattice cell corresponds to its value of solid ratio ($\varepsilon$), which is calculated as the volume covered by the solid particle divided by the total volume of a lattice cell. The colors white ($\varepsilon$ = 0), grey (0 $<$ $\varepsilon$ $<$ 1) and black ($\varepsilon$ = 1) refer to fluid, partially saturated and solid cells, respectively.

The basic principle of the IMB method is to introduce a new collision operator, $\Omega$, depending on the solid ratio, $\varepsilon$. Ideally, the exact value of $\varepsilon$ can be found from a geometrical analysis, but it often requires high computational power. Therefore, a cell decomposition method is adopted, as illustrated in Fig.~\ref{fig:SR}(b). In this method, the partially saturated cells are sub-divided into $n_{sub}^3$ equal-sized sub-cells. An inside-outside algorithm is performed on these sub-cells and $\varepsilon$ is estimated as the number of sub-cells inside the solid boundary (in black) divided by $n_{sub}^3$.

For fluid cells with $\varepsilon$ equal to 0, the normal hydrodynamic collision takes place, and $\Omega$ is taken as the BGK collision operator, $\Omega^f$, as shown on the RHS of Eq.~\eqref{eq:lbe}. For solid cells with $\varepsilon$ equal to 1, a collision operator  proposed by Noble and Torczynski \cite{Noble1998} and based on the concept of non-equilibrium bounce-back \cite{Zou1997} is applied and denoted as $\Omega^s$, which is given by:

\begin{equation}
\Omega_i^s = f_{-i}(\mathbf{x},t)-f_{-i}^{eq}(\rho_f,\mathbf{u}_f)+f_i^{eq}(\rho_f,\mathbf{u}_s)-f_i(\mathbf{x},t),
\label{eq:imb1}
\end{equation}
where $\mathbf{u}_s$ is the macroscopic velocity of solid at the position of the lattice node $\mathbf{x}$. The subscript $-i$ denotes the opposite direction of $i$. The role of the solid collision operator $\Omega^s$ is to ensure a no-slip boundary condition between the fluid phase and the solid phase by setting the PDF, $f_i(\mathbf{x}+\mathbf{c}_i\delta_t, t+\delta_t)$, equal to the EDF, $f_i^{eq}(\rho_f,\mathbf{u}_s)$, plus the bounce-back of the non-equilibrium part in the opposite direction, $f_{-i}(\mathbf{x},t)-f_{-i}^{eq}(\rho_f,\mathbf{u}_f)$.

For partially saturated cells with $\varepsilon$ between 0 and 1, a weighting function, $B$, is used so that it gives:

\begin{equation}
\Omega = B \Omega^s+(1-B) \Omega^f.
\label{eq:partialCollision}
\end{equation}

Following Noble and Torczynski \cite{Noble1998}, the weighting function can be calculated as a function of the relaxation time $\tau$ and the solid ratio $\varepsilon$:

\begin{equation}
B(\varepsilon,\tau) = \frac{\varepsilon(\tau-1/2)}{(1-\varepsilon)+(\tau-1/2)}.
\label{eq:weight}
\end{equation}

Figure~\ref{fig:SR}(c) shows the value of the weighting function $B$ against the solid ratio $\varepsilon$ at four different values of relaxation time: $\tau =$ 0.53, 0.62, 0.8 and 1.0. It can be seen that the $B$ value varies from 0.0 to 1.0 as the solid ratio $\varepsilon$ varies from 0.0 to 1.0. And as the relaxation time $\tau$ increases, the $B$-$\varepsilon$ curve shifts upwards, resulting in a more solid-like behavior for the partially saturated cells. The first term in Eq.~\eqref{eq:partialCollision}, $B\Omega^s$, represents the amount of disturbance to the fluid field due to the presence of solid particles. Therefore, the hydrodynamic force $\mathbf{F}_f$ is the sum of the momentum transfer along all lattice directions at all lattice cells covered by the solid particle (solid and partially saturated lattice cells) with the total number of $n$, which gives:

\begin{equation}
\mathbf{F}_f = \sum \limits_{j=1}^n B_j \sum \limits_{i=0}^{18} \Omega_i^s \mathbf{c}_i.
\label{eq:hydroForce}
\end{equation}

The hydrodynamic torque $\mathbf{T}_f$ is the cross product of the force and the corresponding lever arm, which can be written as:

\begin{equation}
\mathbf{T}_f = \sum \limits_{j=1}^n \left[ B_j (\mathbf{x}_j-\mathbf{x}_s) \times \sum \limits_{i=0}^{18} \Omega_i^s \mathbf{c}_i \right],
\label{eq:hydroTorque}
\end{equation}
where $\mathbf{x}_s$ is the center of mass of the solid particle, and $\mathbf{x}_j$ is the coordinates of the $j$-th lattice cell. The hydrodynamic force and torque calculated from Eq.~\eqref{eq:hydroForce} and Eq.~\eqref{eq:hydroTorque} are back-substituted into Eq.~\eqref{eq:linearVelocity} and Eq.~\eqref{eq:angularVelocity} to update the kinematics and position of each individual solid particle.

\subsection{Coupling scheme}
\label{subsec:coupling}
\begin{figure}[bt]
\centering
\includegraphics{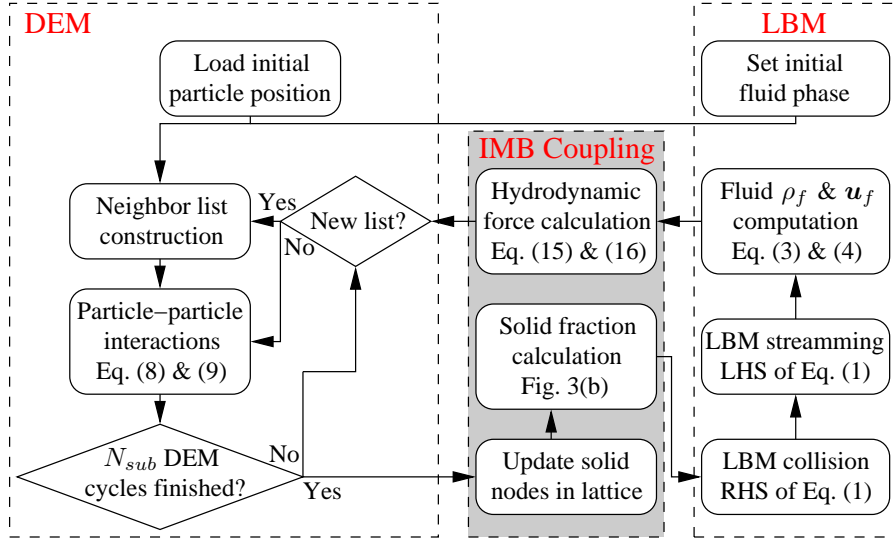}
\caption{Flowchart of the LBM-DEM coupling scheme.}
\label{fig:flowChart}
\end{figure}

Figure~\ref{fig:flowChart} shows the flowchart of the coupling scheme between LBM and DEM. The computing cycle starts with the generation of DEM particles and the initialization of the fluid field, followed by the particle-particle interactions via Eq.~\eqref{eq:normalForce} and Eq.~\eqref{eq:tangentForce}. To achieve a stable DEM simulation, the DEM time step, $\Delta t$, needs to be smaller than a critical value $\Delta t_{cr}$ proportional to  $\sqrt{m/K}$, where $m$ and $K$ are the mass and stiffness of the particles \cite{OSullivan2004}. It is worth noting that the presence of fluid helps to damp the low-frequency elastic waves which can increase the stability of particle simulations, and thereby, a $\Delta t$ value greater than $\Delta t_{cr}$ could be permissible. Nevertheless, it is rather difficult to quantify this stabilization effect. Apart from this, the calculated critical DEM time step $\Delta t_{cr}$ is in general smaller than the time step $\delta_t$ in LBM simulations, especially for problems in the geotechnical field due to the large stiffness of soils and rocks. To synchronize DEM with LBM, $N_{sub}$ DEM sub-cycles are conducted for each step of LBM evolution, so it gives:

\begin{equation}
\Delta t = \frac{\delta_t}{N_{sub}}.
\label{eq:timeStep}
\end{equation}

As a result, during the DEM sub-cycles, the hydrodynamic force $\mathbf{F}_f$ and torque $\mathbf{T}_f$ remain unchanged. It is not an easy task to pre-determine a proper value for $N_{sub}$, which itself is problem dependent. If $N_{sub}$ is too large, it means the degree of coupling between LBM and DEM is weak. Whereas if $N_{sub}$ is too small, a stable DEM simulation might be unattainable. A $N_{sub}$ value smaller than 10 is chosen in \cite{Lomine2013} for the pipe erosion problem. And for the problems simulated in this study with moving particles, we choose $N_{sub} =$ 100 to give smaller DEM time steps for better numerical stability.\

After $N_{sub}$ of DEM sub-cycles, the updated particle positions are mapped on the lattice grid. Then, the lattice cells covered by the solid particles are identified. And if it is a partially saturated cell, the solid ratio $\varepsilon$ is calculated by using the cell decomposition method as shown in Fig.~\ref{fig:SR}(b). According to the state of the lattice cell (fluid, partially saturated or solid), the corresponding collision takes place and the resulting PDFs stream to the neighboring lattice nodes. Based on the redistributed PDFs, the updated fluid density $\rho_f$ and velocity $\textbf{u}_f$ can be calculated from Eq.~\eqref{eq:density} and Eq.~\eqref{eq:velocity}. The data from one LBM cycle are passed back to the IMB coupling module and the hydrodynamic force and torque are calculated based on Eq.~\eqref{eq:hydroForce} and Eq.~\eqref{eq:hydroTorque}, which is further passed back to the DEM module for particle-particle interactions. Up to now, one complete cycle of LBM-DEM simulation is finished and the simulation stops until the specified number of cycles is reached.

\section{Benchmark and numerical issues}
\label{sec:benchmark}

\subsection{Poiseuille flow past a fixed particle}
\label{subsec:poiseFlow}

\subsubsection{Problem description}
\label{subsubsec:poiseFlowDescrip}
\begin{figure}[bt]
\centering
\includegraphics{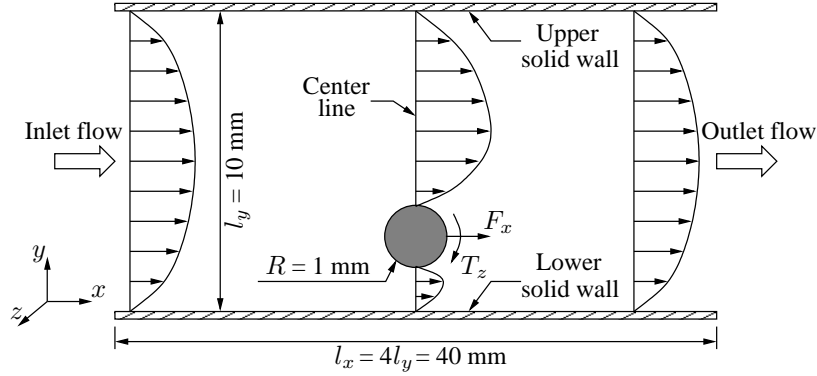}
\caption{2D sketch of a Poiseuille flow past a fixed particle positioned at one-quarter of the distance between the two parallel solid walls.}
\label{fig:poiseFlow}
\end{figure}

The successful coupling between LBM and DEM is first tested against a simple benchmark case: Poiseuille flow past a fixed particle, as shown in Fig.~\ref{fig:poiseFlow}. The hydrodynamic force and torque acting on the particle at steady-state are measured and compared to the available analytical solutions \cite{Happel1983}. Via this simple benchmark case, the influences of two model parameters, referring to the lattice resolution $N$ and the relaxation time $\tau$, which have significant roles in LBM-DEM simulations are discussed in detail. In addition, the least required lattice resolution and the recommendation of selecting the relaxation time are provided.

As shown in Fig.~\ref{fig:poiseFlow}, the fixed particle is positioned between two parallel solid walls with separation of $l_y =$ 10 mm. The radius of the particle is $R =$ 1 mm, which is positioned at half-way in $x$ and $z$ directions and 2.5 mm away from the lower wall measured from the particle center. The upper and lower solid walls in $y$-direction are set to be no-slip boundary conditions, while periodic boundaries are applied in the other two directions. The lengths in $x$ and $z$ directions are large enough with $l_x = l_z = 4l_y =$ 40 mm to make sure that the reflected images of the particle are hydrodynamically decoupled \cite{Clague2001}. The flow is driven from left to right by a body force with an equivalent pressure gradient equal to 2.5E-5 Pa/m. The fluid density ($\rho_f$) and dynamic viscosity ($\nu_f$) are set to be 1000 kg/m$^3$ and 0.001 Pa$\cdot$s. The Reynolds number calculated from the mean flow velocity and the particle diameter is about 4.2E-7, which is small enough to achieve the Stokes flow.

Approximate analytical solutions for the magnitudes of hydrodynamic force $\mathbf{F}_x$ and torque $\mathbf{T}_z$ acting on the particle caused by the drag from the fluid flow are available from Happel and Brenner \cite{Happel1983}, which are given by:

\begin{equation}
F_x = 6\pi \rho_f \nu_f RU \frac{1-1/9(R/l_y)^2}{1-0.6526(R/l_y)+0.316(R/l_y)^3-0.242(R/l_y)^4},
\label{eq:Fx}
\end{equation}

\begin{equation}
T_z = \frac{8}{3}\pi \rho_f \nu_f R^2U \frac{R}{l_y} \left[1+0.0758\left(\frac{R}{l_y}\right)+0.049\left(\frac{R}{l_y}\right)^2\right],
\label{eq:Tz}
\end{equation}
where $U$ is the upstream mean flow velocity.

To compare with Happel and Brenner's estimation as shown in Eq.~\eqref{eq:Fx} and Eq.~\eqref{eq:Tz}, the relative error is defined as follows:

\begin{equation}
\text{Relative error} = \left| \frac{\text{Analytical solution}-\text{Numerical result}}{\text{Analytical solution}} \right|.
\label{eq:error}
\end{equation}

\subsubsection{Effects of the lattice resolution}
\label{subsubsec:latticeReso}
\begin{figure}[bt]
\centering
\includegraphics{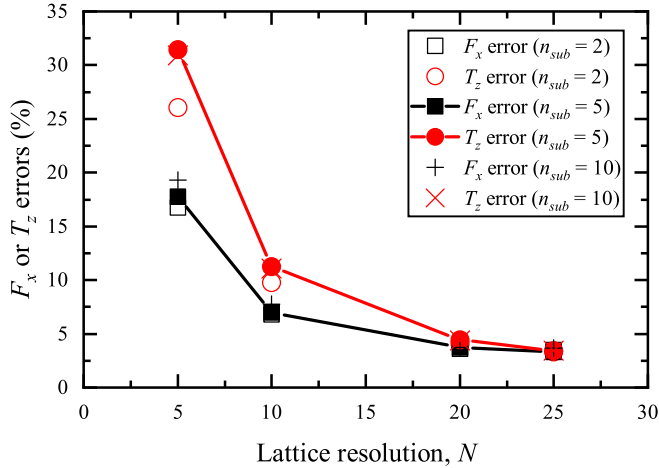}
\caption{Plot of the relative errors for the hydrodynamic force and torque against the lattice resolution $N$ for simulations with various $n_{sub}$ values. The results with $n_{sub} =$ 5 (filled symbols) are connected to show the convergence of the numerical results as $N$ increases. The relaxation time is fixed at $\tau =$ 1.0.}
\label{fig:resolution}
\end{figure}

Similar to many other numerical methods, the accuracy of LBM-DEM simulations highly depends on the spatial (lattice) resolution. In this study, numerical simulations with the lattice resolution, $N$, defined as 5, 10, 20 and 25 are carried out. For each lattice resolution, the number of sub-slice $n_{sub}$ varies between 2, 5 and 10 for the calculation of the solid ratio $\varepsilon$. The relative force and torque errors for different $n_{sub}$ values are plotted against the lattice resolution in Fig.~\ref{fig:resolution}. For these simulations, the relaxation time is fixed at $\tau =$ 1.0.

When the lattice resolution is low, with $N =$ 5, the force and torque errors can be as high as 17.8\% and 31.4\%, when $n_{sub} =$ 5. As $N$ increases from 5 to 20, both the calculated hydrodynamic force and torque errors decrease rapidly. At $N =$ 20, the LBM-DEM model is able to produce results within 4.5\% of errors compared to Happel and Brenner's estimation. Further increase of lattice resolution from $N =$ 20 to $N =$ 25 only yields a minimal improvement in accuracy. Hence, a spatial resolution of at least 20 number of lattice cells across one particle diameter is recommended to achieve reasonably accurate coupled LBM-DEM simulations using the IMB method. This conclusion also stands for cases with multiple particles, see Section~\ref{subsec:flowPack} and Section~\ref{subsec:couette}, since both the force and the torque here are local measurements. The recommended lattice resolution here is much higher than the current computational practice with $N =$ 10, which agrees with Rettinger's recent findings by looking at the settling of a single sphere \cite{Rettinger2017}.

The fluidity of the partially saturated cells in the LBM-DEM model is determined by their solid ratios. And indeed, the accuracy of solid ratio calculation does affect the LBM-DEM results significantly when the lattice resolution is low. However, if the lattice resolution is sufficiently high, for example $N =$ 20, the influence from $n_{sub}$ is negligible. Therefore, for the simulations presented in the rest of this paper, we use 5 sub-slices or 125 sub-cells to calculate the solid ratio for each partially saturated cell, which should be accurate enough. And it will also be shown in Section~\ref{subsubsec:cost} that $n_{sub}$, within the range of consideration, is not the major factor affecting the computational time.

\subsubsection{Effects of the relaxation time}
\label{subsubsec:relaxTime}
\begin{figure}[bt]
\centering
\includegraphics{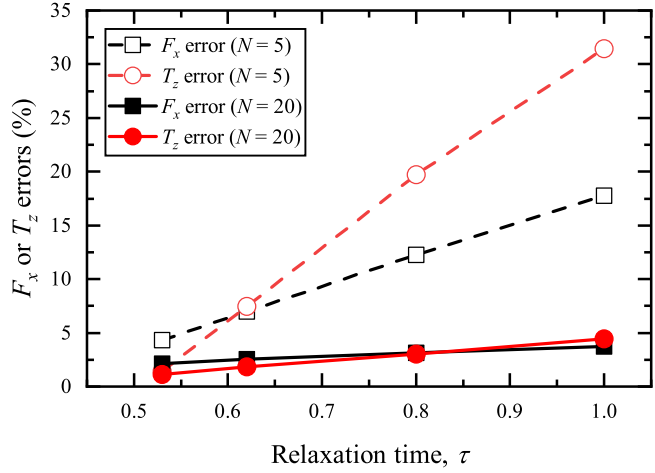}
\caption{Plot of the relative force and torque errors against the relaxation time $\tau$ at lattice resolutions $N =$ 5 and $N =$ 20.}
\label{fig:relaxationTime}
\end{figure}

The relaxation time $\tau$ physically determines how fast the PDFs recover the current EDFs, as shown in Eq.~\eqref{eq:lbe}. Previous study has already revealed that the BGK (or single-relaxation-time) model adopted in our work may lead to inaccurate no-slip boundary locations \cite{Luo2011}. To investigate the influence of the relaxation time $\tau$ on the coupling between LBM and DEM, numerical simulations with $\tau$ equal to 0.53, 0.62, 0.8 and 1.0 are carried out with the fluid viscosity being unchanged. The relative force and torque errors are plotted against the relaxation time in Fig.~\ref{fig:relaxationTime} at a low ($N =$ 5) and a high ($N =$ 20) lattice resolution.

Interestingly, it can be seen that as $\tau$ increases, both the relative force and torque errors increase roughly in a linear way. The dependence of the LBM-DEM results on the relaxation time is much more significant when the lattice resolution is low at $N =$ 5. The relaxation time is able to affect the accuracy of the LBM-DEM model in several different ways, which can be divided into two categories: one belongs to the fluid solution only, and the other comes from the fluid-particle interactions.

First, for a constant fluid viscosity $\nu_f$ and a fixed lattice spacing $\delta_x$, the LBM time step $\delta_t$ decreases as the relaxation time $\tau$ decreases, according to Eq.~\eqref{eq:viscosity}. Therefore, when the fluid velocity is normalized by the term $\delta_x/\delta_t$, the resultant fluid velocity in lattice units is also reduced. In this way, the Mach number drops as $\tau$ decreases, resulting in smaller compressibility errors for the fluid solution.

\begin{figure}[bt]
\centering
\includegraphics{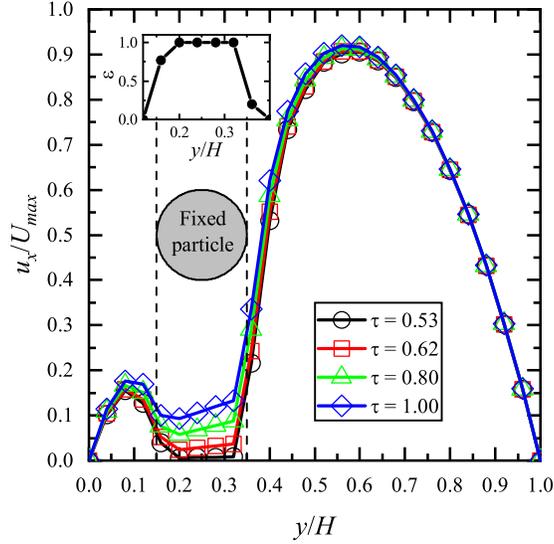}
\caption{Profiles of the normalized streamwise velocity at the section in $y$-direction going through the center of the particle from LBM-DEM simulations with various relaxation times: $\tau =$ 0.53, 0.62, 0.8 and 1.0. The lattice resolution is $N =$ 5.}
\label{fig:velProfile}
\end{figure}

When the fluid is coupled with the particles, the relaxation time has a significant influence on the no-slip boundary conditions \cite{Luo2011}. In order to examine this issue, the streamwise velocity profiles at the center line, as shown in Fig.~\ref{fig:poiseFlow}, are plotted in Fig.~\ref{fig:velProfile} for simulations with various relaxation times. The lattice resolution is $N =$ 5 since the influence of the relaxation time becomes more obvious when the lattice resolution is low according to Fig.~\ref{fig:relaxationTime}. The streamwise velocity $u_x$ is normalized by the theoretical maximum flow velocity $U_{max}$ in the absence of the particle. The inserted figure shows the distribution of the solid ratio.

As shown in Fig.~\ref{fig:velProfile}, when $\tau$ is large and close to 1.0, the fluid momentum is able to diffuse into the particles, producing non-zero flow velocities at the solid cells (with $\varepsilon = 1$). As $\tau$ decreases from 1.0 to 0.53, the flow velocities at the solid cells decrease to nearly zero values, indicating a highly improved no-slip boundary condition. As a result, the involved hydrodynamic interactions via momentum exchange between the fluid and the solid particle are also better described as $\tau$ decreases. Due to this diffusion effect of fluid momentum, the fluid field can only \emph{feel} a \emph{transparent} particle, resulting into the underestimated drag force especially when $\tau$ is large, which agrees with the results in Fig.~\ref{fig:relaxationTime}.

The relaxation time can also affect the LBM-DEM results via the weighting function $B(\varepsilon, \tau)$, as shown in Fig.~\ref{fig:SR}(c). A more solid-like behavior is observed for the partially saturated cells as $\tau$ increases. However, the currently adopted weighting function is still not able to adequately compensate for the weakened fluid-particle interaction due to the diffusion effect. In future works, a super-linear relationship between $B$ and $\varepsilon$ can be proposed to potentially increase the accuracy of the LBM-DEM model when $\tau$ increases.

\subsubsection{Computational cost}
\label{subsubsec:cost}
The calculated hydrodynamic force and torque from several selected LBM-DEM simulations and the relative errors compared to Happel and Brenner's estimation are listed in Table~\ref{tab:forceTorque}. All cases are simulated with two compute nodes, each of which is equipped with two 10-core Intel Xeon E5-2600 v3 processors and 96 GB physical memory. The total simulation time, $T$, is presented at the last column of Table~\ref{tab:forceTorque} in minutes.

\begin{table}[bt]
\caption{Comparison between the LBM-DEM results and Happel and Brenner's estimation \cite{Happel1983} in terms of the hydrodynamic force and torque acting on a fixed particle in Poiseuille flow. The analytical solutions for the hydrodynamic force and torque are 5.7628E-12 N and 8.1537E-16 N$\cdot$m, respectively.}
\begin{tabular}{l l l l l l l l l}
\hline
$N$ & $n_{sub}$ & $\tau$ & $F_x$ (E-12 N) & $T_z$ (E-16 N$\cdot$m) & Error $F_x$ (\%) & Error $T_z$ (\%) & $T$ (min) \\
\hline
\textbf{5}  & 5  & 1.0  & 4.7375 & 5.5904 & 17.7910 & 31.4374 & 0.58   \\
\textbf{10} & 5  & 1.0  & 5.3606 & 7.2359 & 6.9793  & 11.2560 & 8.53   \\
\textbf{20} & 5  & 1.0  & 5.5462 & 7.7928 & 3.7595  & 4.4260  & 182.52 \\
\textbf{25} & 5  & 1.0  & 5.5709 & 7.8732 & 3.3305  & 3.4398  & 482.77 \\
20 & \textbf{2}  & 1.0  & 5.5553 & 7.8202 & 3.6016  & 4.0905  & 184.05 \\
20 & \textbf{4}  & 1.0  & 5.5484 & 7.8001 & 3.7203  & 4.3372  & 173.90 \\
20 & \textbf{10} & 1.0  & 5.5476 & 7.7986 & 3.7343  & 4.3546  & 180.15 \\
20 & 5  & \textbf{0.8}  & 5.5829 & 7.9051 & 3.1223  & 3.0488  & 296.75 \\
20 & 5  & \textbf{0.62} & 5.6166 & 8.0031 & 2.5379  & 1.8471  & 721.40 \\
20 & 5  & \textbf{0.53} & 5.6385 & 8.0608 & 2.1564  & 1.1390  & 4312.90\\
\hline
\end{tabular}
\label{tab:forceTorque}
\end{table}

Note that the simulation time is directly related to the performance of the computer. And the influence of each model parameter on the computational cost depends on the complexity of the problem, for example, the number of DEM particles involved. However, we can still identify two parameters that play a major role in the computational demand, including the lattice resolution $N$ and the relaxation time $\tau$. Generally speaking, the computational cost increases rapidly as $N$ increases and $\tau$ decreases. In return, the accuracy of the LBM-DEM model is usually improved.

\subsection{Particle settling in an ambient fluid}
\label{subsec:particleSettle}

\subsubsection{Problem description}
\label{subsubsec:particleSettleDescrip}
\begin{figure}[bt]
\centering
\includegraphics{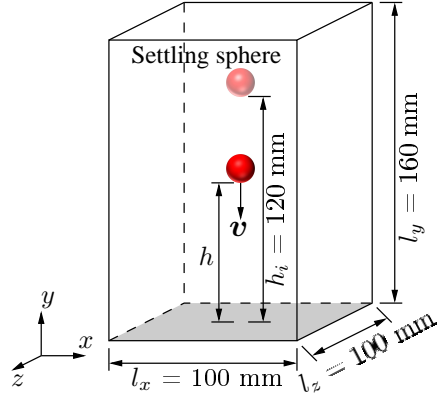}
\caption{Sketch of a single heavy particle settling in an ambient fluid.}
\label{fig:settle}
\end{figure}

In order to test the sub-cycling scheme and the additional numerical error caused by particle moving across multiple lattice cells, the case of a single heavy particle settling in an ambient fluid is simulated, as shown in Fig.~\ref{fig:settle}. The numerical test setup is the same as the physical experiment conducted in \cite{Cate2002}. A particle with diameter $d_p =$ 15 mm is released at an initial height $h_0 =$ 120 mm in a container filled with fluid. The initial velocity is set to be zero. The container has a dimension of 100 mm, 160 mm, and 100 mm in $x$, $y$ and $z$ directions, respectively. During the sedimentation of the particle, the settling velocity $\textbf{v}$ and the distance to the bottom of the container $h$ are recorded.

We have repeated the same experiments referenced in \cite{Cate2002} with four different particle Reynolds numbers: Re = 1.5 (E1), 4.1 (E2), 11.6 (E3) and 31.9 (E4), based on the terminal velocity of the particle, by varying the fluid density and fluid viscosity. The lattice resolution and the relaxation time are correspondingly set to be 20 and 0.62 for all four cases, which gives the LBM time steps equal to 5.85E-5 s, 1.02E-4 s, 1.91E-4 s and 3.72E-4 s for E1 to E4, respectively. The solid walls in all directions are set as no-slip boundary conditions using the bounce-back method \cite{Luo2011}.

\subsubsection{Results and discussion}
\label{subsubsec:results1}
\begin{figure}[bt]
\centering
\includegraphics{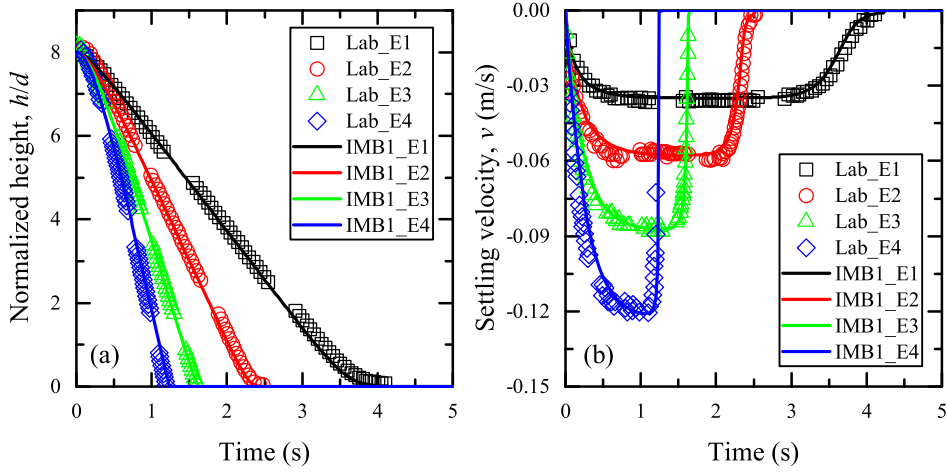}
\caption{Comparison between the simulated and measured results regarding to (a) the particle settling trajectory and (b) the particle settling velocity evolution. In total, four different cases are simulated, including E1 (Re = 1.5), E2 (Re = 4.1), E3 (Re = 11.6) and E4 (Re = 31.9).}
\label{fig:trajVel}
\end{figure}

Fig.~\ref{fig:trajVel} shows the comparison between the calculated numerical results and the measured experimental data in terms of the particle trajectory and the evolution of settling velocity. It can be seen that for the case E1 with a low Reynolds number and high fluid viscosity, the settling velocity first increases until the terminal velocity is reached, at which the weight of the particle is balanced by the buoyancy force and the drag force. After that, the particle velocity decreases slowly as it approaches the bottom solid wall due to the additional force produced by lubrication effects \cite{Batchelor2000}. For the case E4 with a high Reynolds number and small fluid viscosity, the particle settles with a much higher acceleration and velocity, and then decelerates quickly. A settling period with stable terminal velocity is barely observed.

Note that no artificial lubrication force is incorporated in our LBM-DEM model, the accurate description of the particle movement when it approaches the bottom solid wall highlights the sub-grid scale resolution of the IMB method. Unlike the ME method \cite{Ladd1994}, a layer of fluid is solved by the partially saturated cells when $h < \delta_x$. All in all, the agreement between the numerical results and the experimental data serves as the evidence for the reliability of the proposed coupling scheme for problems with moving particles in an ambient fluid.

\subsection{Flow through densely packed particles}
\label{subsec:flowPack}

\subsubsection{Problem description}
\label{subsubsec:flowPackDescrip}
\begin{figure}[bt]
\centering
\includegraphics{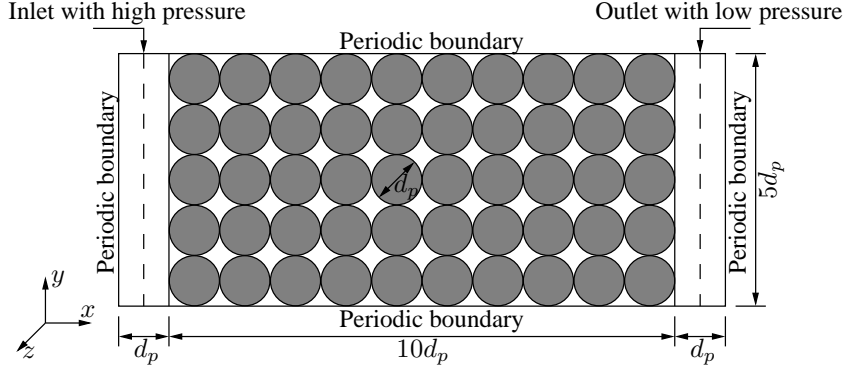}
\caption{Numerical setup of flow through densely packed particles in a cubic arrangement.}
\label{fig:porous}
\end{figure}

In order to examine the accuracy of the LBM-DEM model at the inertial flow regime, a flow through a densely packed granular medium is simulated, as shown in Fig~\ref{fig:porous}. The granular medium consists of monodispersed particles with diameter $d_p =$ 1 mm packed in a simple cubic arrangement. The fluid has a density $\rho_f =$ 1000 kg/m$^3$ and a dynamic viscosity $\mu_f =$ 0.001 Pa$\cdot$s. The granular medium has a dimension of 10$d_p$, 5$d_p$ and 5$d_p$ in $x$, $y$ and $z$ directions, respectively. An additional one $d_p$ of spacing is left at the inlet and outlet for the development of inflow and outflow. Periodic boundaries are defined in all directions. The flow is driven from left to right by ten different pressure differences ($\Delta P$), including 1 Pa, 5 Pa, 10 Pa, 20 Pa, 50 Pa, 100 Pa, 200 Pa, 400 Pa, 800 Pa and 1600 Pa. Again, the lattice resolution is set to be $N =$ 20. As $\Delta P$ increases, the fluid velocity also increases, therefore a smaller relaxation time is required to reduce the compressibility error. In order to keep the maximum fluid density variation below 0.6\%, the relaxation time is gradually reduced by following the order of 0.58, 0.53, 0.524, 0.516, 0.51, 0.508, 0.505, 0.503, 0.502 and 0.501 with the increase of $\Delta P$. All simulations last for 5 s, which is long enough to allow the flow to be fully developed.

The total pressure loss can be described by the well-known Ergun equation \cite{Ergun1952} which is composed of two terms: a viscous loss proportional to the fluid velocity and an inertia loss proportional to the square of the fluid velocity, given by:

\begin{equation}
\frac{\Delta P}{L} = \frac{150\mu_f}{d_p^2} \frac{(1-n)^2}{n^3}U+\frac{1.75\rho_f}{d_p} \frac{(1-n)}{n^3}U^2,
\label{eq:ergun}
\end{equation}
where $L$ is the length of the granular medium and equal to 10$d_p$. The porosity is denoted as $n$, which is equal to 0.4764. The superficial fluid velocity is denoted as $U$.

Following Niven \cite{Niven2002}, Eq.~\eqref{eq:ergun} can be rewritten in the form of:

\begin{equation}
f^* = \frac{\Delta P}{L} \frac{d_p}{\rho_f U^2} \frac{n^3}{(1-n)} = \frac{150}{\text{Re}_p^*}+1.75,
\label{eq:ergun2}
\end{equation}
where $f^*$ is known as the packed bed friction factor. $\text{Re}_p^*$ is a modified particle Reynolds number based on the interstitial fluid velocity, which is given by:

\begin{equation}
\text{Re}_p^* = \frac{\rho_f d_p U}{\mu_f(1-n)}.
\label{eq:modifiedRe}
\end{equation}

\subsubsection{Results and discussion}
\label{subsubsec:results2}
Previous work \cite{Han2013} suggested that the LBM-DEM model produced $f^*$ smaller than the Ergun equation with a rapid increase of numerical error as $\text{Re}_p^*$ was greater than 40. In addition, the numerical results tended to collapse with the predictions of the Ergun equation if a higher lattice resolution was adopted. A similar result was observed in \cite{Rettinger2017} that the lattice resolution needed to be increased to accurately solve the fluid-particle interactions in inertial flows. It becomes reasonable to speculate that the underestimation from LBM-DEM models could be due to the lack of resolution to solve the small-scale eddies when the Reynolds number is high. Therefore, we borrow the idea of large eddy simulation \cite{Hou1996}, and an SGS-Smagorinsky turbulence model is incorporated into the LBM simulation.

\begin{figure}[bt]
\centering
\includegraphics{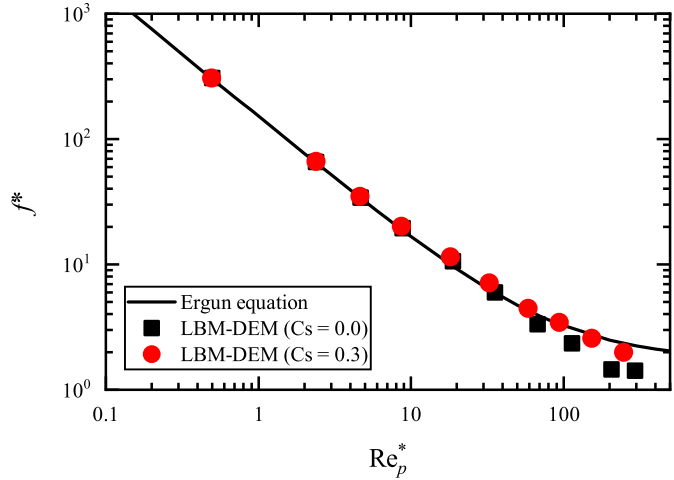}
\caption{Comparison between the LBM-DEM results and the Ergun equation in terms of the packed bed friction factor $f^*$ at various modified particle Reynolds numbers $\text{Re}_p^*$.}
\label{fig:ergun}
\end{figure}

LBM-DEM simulations with and without the turbulence model are conducted, the resultant friction factor $f^*$ is plotted against the modified particle Reynolds number $\text{Re}_p^*$ in Fig.~\ref{fig:ergun}. The empirical Ergun equation, that is Eq.~\eqref{eq:ergun2}, is also plotted for comparison. It can be seen that when $\text{Re}_p^*$ is small, the LBM-DEM results are in good agreement with the Ergun equation. And at small $\text{Re}_p^*$, the incorporation of the turbulence model results in negligible effects due to the near-zero eddy viscosity \cite{Hou1996}.

However, as $\text{Re}_p^*$ increases, the small-scale eddies developed inside the granular medium cannot be well captured by the LBM method, leading to underestimated friction factors. Fig.~\ref{fig:ergun} shows that the incorporation of the turbulence model with the Smagorinsky constant $C_s =$ 0.3 can successfully bring the LBM-DEM results back to the trend of the Ergun equation. It is also worth mentioning that the proper value of $C_s$ is problem dependent and here it is determined by trial and error. It can be seen that at a moderate $\text{Re}_p^*$ between 10 and 100, the LBM-DEM simulations, together with the turbulence model, overestimate the Ergun equation. For these cases, a smaller $C_s$ value should be used.

All in all, the LBM-DEM simulations using the IMB method can well match the Ergun equation within a wide range of $\text{Re}_p^*$ from $\mathcal{O}(10^{-3})$ to $\mathcal{O}(10^2)$, when a turbulence model is incorporated, which is quite encouraging based on the fact that the LBM-DEM model only resolves the local fluid-particle interactions at the pore-scale, while the Ergun equation describes the overall resistance on the fluid field from the whole porous structure. In another word, the macroscopic behavior is automatically recovered from the microscopic fluid-particle interactions.

\subsection{Couette flow of particle suspensions}
\label{subsec:couette}

\subsubsection{Problem description}
\label{subsubsec:couetteDescrip}
In a practical fluid-particle interaction problem, such as debris flow, the dynamic of the fluid-particle mixture is governed by the particle-particle interactions either by direct contact or via the interstitial fluid. To highlight the capability of the proposed LBM-DEM model in capturing the complex particle-fluid-particle interactions, a problem involving multiple and movable particles submerged in a fluid is simulated. Fig.~\ref{fig:couette} shows a concentrated suspension of neutrally buoyant and monodispersed particles with diameter $d_p =$ 1 mm. The fluid density and dynamic viscosity are set to be $\rho_f =$ 1000 kg/m$^3$ and $\mu_f =$ 0.001 Pa$\cdot$s, respectively. The simulation domain has a size equal to 10$d_p$, 10$d_p$ and 5$d_p$ in the $x$, $y$, and $z$ directions, respectively. The flow is driven by moving the top and bottom solid walls to the right and to the left with a constant velocity $u_w =$ 0.001 m/s. Periodic boundaries are defined in the $x$ and $z$ directions. In this way, an average shear rate, $\dot{\gamma}$, in the fluid can be calculated, as:

\begin{figure}[bt]
\centering
\includegraphics{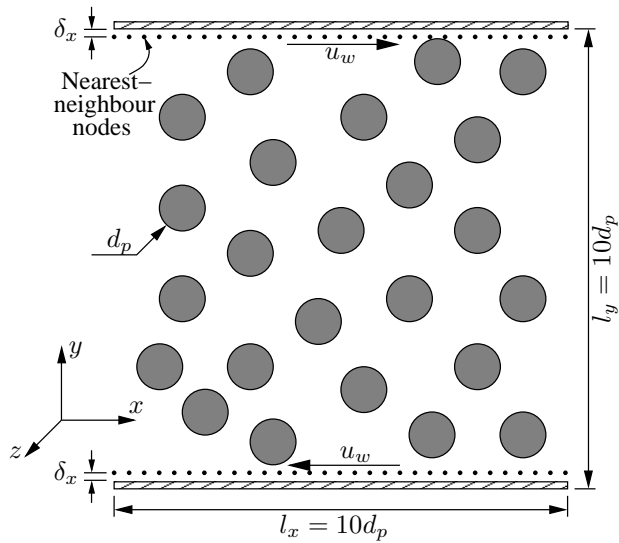}
\caption{2D sketch of a concentrated suspension undergoing planar Couette flow.}
\label{fig:couette}
\end{figure}

\begin{equation}
\dot{\gamma} = \frac{2u_w}{H},
\label{eq:shearRate}
\end{equation}
where $H$ is the distance between the two parallel solid walls. The wall shear stress, $\tau_w$, is given by:

\begin{equation}
\tau_w = \mu_f \dot{\gamma}.
\label{eq:shearStress}
\end{equation}

Equation~\eqref{eq:shearStress} is usually applied to measure the fluid viscosity via rheometers. And for a pure Newtonian fluid, $\mu_f$ remains constant, independent on the magnitude of shear rate. However, if there are particles suspended in the fluid, the rheology of the mixture becomes different from that of the pure fluid. Generally speaking, the apparent viscosity of the particle suspension $\mu_f^*$ increases as the solid volume fraction $\phi_p$ increases. The relationship between $\mu_f^*$ and $\phi_p$ is first described by Einstein \cite{Mooney1951}, written as:

\begin{equation}
\mu_f^*(\phi_p) = \mu_f(1+2.5\phi_p).
\label{eq:Einstein}
\end{equation}

The Einstein's viscosity equation is deduced with the assumption of negligible particle-particle interactions. Therefore, Eq.~\eqref{eq:Einstein} is only valid for extremely dilute systems. Following Einstein, researchers have spent huge efforts, trying to extend Einstein's viscosity equation to suspensions with finite concentrations. One of the most popular ones is the classic work from Mooney \cite{Mooney1951}. For a suspension of monodispersed particles:

\begin{equation}
\mu_f^*(\phi_p) = \mu_f \exp\left(\frac{2.5\phi_p}{1-k\phi_p}\right),
\label{eq:Mooney}
\end{equation}
where $k$ is the self-crowding factor. If mechanical interlocking takes place at the densest possible state, which is the face-centered cubic packing with $\phi_p =$ 0.74, the apparent viscosity becomes infinitely large. Then, $k$ takes the value of 1.35.

In this study, six simulations have been carried out with the solid volume fraction $\phi_p =$ 0.0, 0.0199, 0.0503, 0.0953, 0.1414 and 0.1571. The particles are created with zero initial velocity. Again, the fluid-particle interaction is solved with a lattice resolution $N =$ 20. The relaxation time is set to be 0.8. All simulations last for 14 seconds so that a steady state can be obtained.

\subsubsection{Results and discussion}
\label{subsubsec:results3}
\begin{figure}[bt]
\centering
\includegraphics{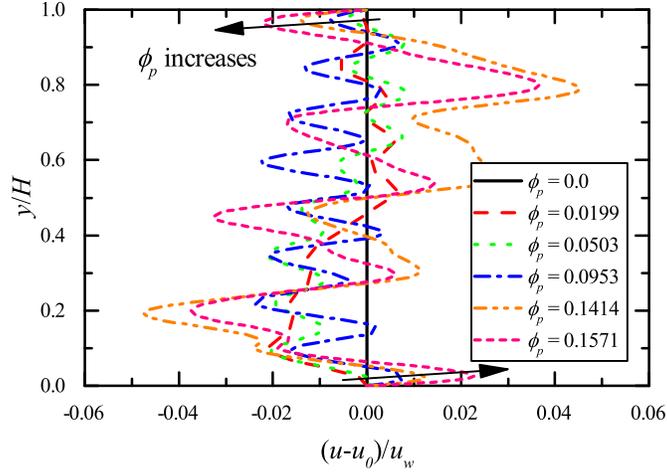}
\caption{Spatial distribution of the normalized velocity difference, $(u-u_0)/u_w$, across the planar Couette flow of suspensions with various solid volume fractions.}
\label{fig:velDiff}
\end{figure}

Figure~\ref{fig:velDiff} shows the spatial distribution of the velocity difference $(u-u_0)$ normalized by the wall velocity $u_w$, where $u$ is the calculated flow velocity averaged in the $x$ and $z$ directions and $u_0$ is the theoretical linear profile for the case of pure fluid. First of all, when $\phi_p =$ 0.0, the theoretical linear distribution of the Couette flow velocity is well recovered by the LBM solver, which is evidenced by the zero $(u-u_0)/u_w$ value as $y/H$ goes from 0.0 to 1.0. As $\phi_p$ increases, the amount of fluctuation also increases, showing a non-Newtonian behavior. Particularly, the concentrated suspension with a higher solid volume fraction $\phi_p$ has a steeper velocity gradient close to the top and the bottom solid walls.

\begin{figure}[bt]
\centering
\includegraphics{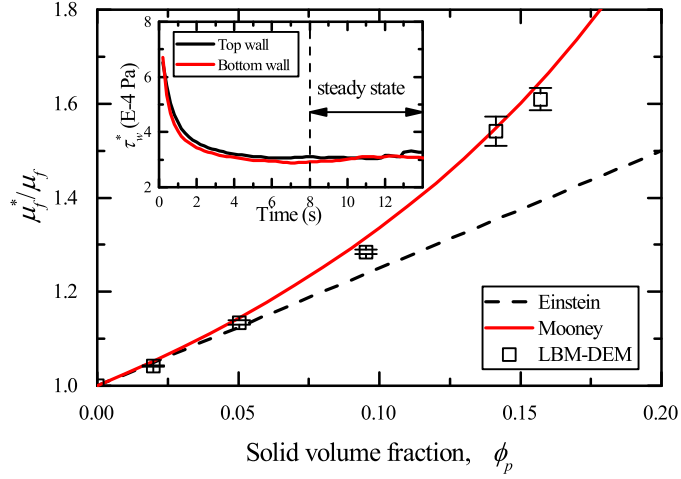}
\caption{Comparison between the LBM-DEM numerical results with the Einstein's and Mooney's viscosity equation in terms of the relative viscosity ratio, $\mu_f^*/\mu_f$, at various solid volume fractions. The error bar shows the standard deviation of the numerical results. The inserted figure shows a typical evolution of the apparent shear stresses at the top and the bottom solid walls since the start of the simulation, when $\phi_p =$ 0.1414. A steady state is reached after 8 seconds.}
\label{fig:viscosity}
\end{figure}

The strain rate at the solid boundaries are calculated from the wall velocity $u_w$ and the flow velocity at the nearest-neighbor fluid node $u_{\alpha,\beta}$, see Fig.~\ref{fig:couette}, where the subscripts $\alpha$ and $\beta$ are the spatial indices in $x$ and $z$ directions. Make $n_{\alpha}$ and $n_{\beta}$ be the number of lattice nodes in the $x$ and $z$ directions, respectively, the apparent wall shear stress $\tau_w^*$ is given by:

\begin{equation}
\tau_w^* = \mu_f^* \dot{\gamma} = \frac{\mu_f \sum_1^{n_{\alpha}} \sum_1^{n_{\beta}} |u_w-u_{\alpha,\beta}|/\delta_x}{n_{\alpha} n_{\beta}},
\label{eq:wallShear}
\end{equation}
where $\mu_f^*$ is the apparent dynamic viscosity. The inserted figure of Fig.~\ref{fig:viscosity} show a typical development of the apparent wall shear stresses $\tau_w^*$ when the solid volume fraction is $\phi_p =$ 0.1414. At the early stage, additional momentum is required to bring the stationary particles into motion. The reaction forces acting on the fluid field result in high shear stresses at the solid walls. As time goes by, the particles gradually accelerate and the wall shear stresses gradually decrease until a steady state is reached, which is about 8 s after the start of simulation. In order to calculate the apparent fluid viscosity using Eq.~\eqref{eq:wallShear}, $\tau_w^*$ take the averaged wall shear stress at the top and the bottom solid walls, which is further averaged against the time from 8 s to 14 s.

Fig.~\ref{fig:viscosity} shows the variation of the relative viscosity ratio, $\mu_f^*/\mu_f$, with the solid volume fraction. The error bar indicates the standard deviation of the LBM-DEM result due to temporal variations. The Einstein's and Mooney's viscosity equations are also plotted for comparison. It can be seen that the Einstein's equation only fits the data when $\phi_p$ is smaller than 0.02. While the LBM-DEM result well agrees with the Mooney's equation for the whole range of $\phi_p$ tested in this study.

\section{Concluding remarks}
\label{sec:conclusions}
\begin{figure}[bt]
\centering
\includegraphics{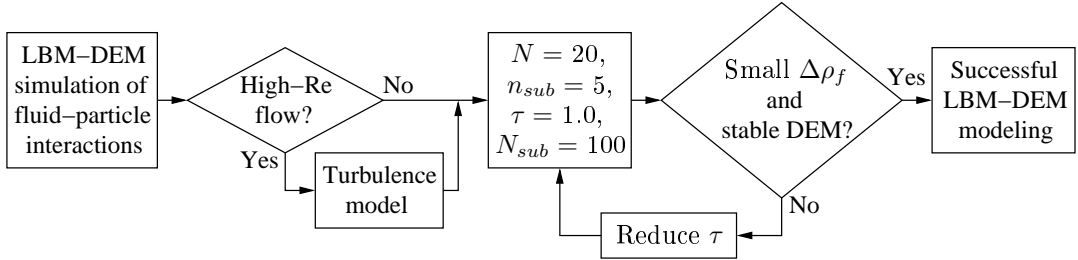}
\caption{Procedures to build up an accurate, efficient and stable LBM-DEM model.}
\label{fig:guideline}
\end{figure}

We would like to conclude this paper by providing a general guideline for setting up an accurate, efficient and stable LBM-DEM model, as shown in Fig.~\ref{fig:guideline}.

First of all, the accuracy of the LBM-DEM model is highly dependent on the flow regime of the problem under consideration. For high-Reynolds number flows, it becomes necessary to fully resolve the small-scale eddies, which requires a higher spatial resolution. However, the increase of spatial resolution may lead into unaffordable computational demand, especially for 3D simulations involving hundreds and thousands of particles. Alternatively, the effects from unresolved small eddies can be well captured by incorporating a turbulence model, such as, the simple SGS-Smagorinsky model used in this study. By this way, the fluid-particle interactions can be resolved over a wide range of Reynolds numbers in and accurate and efficient manner.

Second, it is recommended that a resolution of 20 lattice cells across one particle diameter ($N =$ 20) shall be used so that a highly accurate 3D LBM-DEM simulation can be achieved. This recommendation is based on the calculation of the fluid drag on a single particle at the Stokes regime, in which turbulence does not play a role. Combined with a valid turbulence model, the fluid-particle interactions in flows with a wide range of Reynolds numbers can be well captured in an accurate and efficient manner. Since the LBM-DEM model is able to fully resolve the momentum exchange between the fluid and each individual particle locally, the same lattice resolution can be applied for systems involving multiple particles. It is also found that the accuracy of solid ratio calculation for partially saturated cells does not have a significant effect on the LBM-DEM results, as long as an adequate lattice resolution is used. A resolution of 5 sub-slices ($n_{sub} =$ 5) is high enough to offer a good estimation when the cell decomposition method is adopted.

Third, the relaxation time $\tau$ shall be chosen depending on the Mach number. When the fluid velocity is large, the Mach number is large, a smaller $\tau$ value is required to keep the compressibility error small. An initial relaxation time equal to 1.0 can be taken for the first trial of the LBM-DEM simulation. If the fluid density variation is too large, simulations with smaller $\tau$ values, larger than the lower limit 0.5, need to be conducted. Besides, as $\tau$ decreases, the fluid-particle interaction is better resolved and a higher degree of coupling between LBM and DEM becomes possible due to the reduced LBM time step. It enhances the stability of DEM simulations. However, a decrease in the relaxation time also comes with higher computational effort.

Forth, the number of sub-cycling $N_{sub}$ is suggested to be 100 at maximum so that a reasonably strong coupling between LBM and DEM can be achieved.

The above-mentioned procedure will allow users to effectively build-up a LBM-DEM model with high accuracy.

\section*{References}

\bibliography{mybibfile}

\end{document}